\documentstyle[preprint,revtex]{aps}

\begin{document}
\draft

\begin{title}
Influence of Quasi-Bi-Stripe Charge Order on the Resistivity \\
and Magnetism in a Bilayer Manganite La$_{2-2x}$Sr$_{1+2x}$Mn$_{2}$O$_{7}$
% ($0.30 \le x \le 0.50$)
\end{title}

\begin{center}
\author{M. Kubota$^*$, Y. Oohara, H. Yoshizawa}
\begin{instit}
Neutron Scattering Laboratory, I.S.S.P., University of Tokyo, Tokai, Ibaraki, 319-1106
\end{instit}
\author{H. Fujioka, K.  Shimizu, K. Hirota} 
\begin{instit}
CREST, Department of Physics, Tohoku University, Aoba-ku, Sendai, 980-8578\\
\end{instit}
\author{Y. Moritomo}
\begin{instit}
Center for Integrated Research in Science and Engineering, Nagoya University, Nagoya, 464-8601
\end{instit}
\author{ Y. Endoh}
\begin{instit}
Institute for Materials Research, Tohoku University, Aoba-ku, Sendai 980-8577\\
\end{instit}

\end{center}

\date{\today}

%\twocolumn[\hsize\textwidth\columnwidth\hsize\csname @twocolumnfalse\endcsname
%\maketitle

\begin{abstract}
The charge ordering in the bilayer manganite system La$_{2-2x}$Sr$_{1+2x}$Mn$_{2}$O$_{7}$ with $0.30 \le x \le 0.50$ has been studied by neutron diffraction.  The charge order is characterized by the propagation vector parallel to the [1 0 0] direction ( MnO$_2$ direction), but the correlation length is short-ranged and extremely anisotropic, being $\sim 0.02a^{*}$ and $\sim 0.2a^{*}$ parallel and perpendicular to the modulation direction, respectively. The observed charge order can be viewed as a quasi-bi-stripe order, and accounts well for the $x$ dependence of the resistivity.  The quasi-bistripe order is stable within the ferromagnetic (FM) MnO$_2$ layers in the A-type antiferromagnetic order, but is instabilized by the 3 dimensional FM order.
\end{abstract}

%\kword {bilayer manganite La$_{2-2x}$Sr$_{1+2x}$Mn$_{2}$O$_{7}$, 
%colossal magnetoresistance (CMR), charge ordering, 
%stripe order,  $e_{g}$ orbital, low-dimensionality}
%]

%\sloppy
%\maketitle

It has been recently demonstrated that an A-type antiferromagnetic (AFM) state in the manganites near the 50\% hole concentration shows the metallic conductivity\cite{kaw97}.  The A-type AFM spin order consists of antiferromagnetically stacked ferromagnetic (FM) layers, but the metallic A-type AFM state should be strictly differentiated from the insulating A-type AFM state which has been observed in the non-doped compound LaMnO$_{3}$ with strong Jahn-Teller distortion.  The essence of the former metallic A-type state is the orbital ordering of the $d_{x^2-y^2}$ orbitals within FM MnO$_{2}$ layers.  A series of studies of Nd$_{1-x}$Sr$_{x}$MnO$_{3}$\cite{kaw97,kuwahara99,mor98} has demonstrated that the anisotropy of the $d_{x^2-y^2}$-type orbitals causes anisotropic behavior of the resistivity as well as the exchange interactions with respect to the FM MnO$_{2}$ layers.

We have recently reported that the {\it A-type AFM phase} also appears in a two-dimensional (2D) bilayer system La$_{2-2x}$Sr$_{1+2x}$Mn$_{2}$O$_{7}$ for $0.40 \leq x \leq 0.50$ \cite{hirota98,kubota99a,kubota99b}.  For instance, the $x=0.45$ sample exhibits an A-type AFM Bragg reflection below $T_{\rm N} \sim$ 200~K as shown in Fig. 1(a), but the ferromagnetic component appears below $T_{\rm C} \sim 95$ K, resulting in a canted AFM state at low temperatures. The A-type AFM phase in the $x=0.45$ sample is, however, anomalous in many respects.  As seen in Fig. 1(a), the $T$ dependence of the intensity of $(0 0 3)$ does not follow a conventional Brillouin function, but exhibits a logarithmic behavior for $T_{\rm C} \le T \le T_{\rm N}$.  This suggests strong spin fluctuations which suppress the A-type AFM long range order.

Corresponding to the anomalous $T$ dependence of the Bragg intensity, the slope of the H/M exhibits a clear change around $T_{\rm N}$ (Fig. 1(b)).  The anomaly in the susceptibility around $\sim 250$ K was pointed out in early magnetization measurements by several works, but its origin was not elucidated.\cite{Mitchell97,kim97}  In our previous study,\cite{hirota98} it was identified as the onset of the anomalous A-type AFM ordering.  Since the structural analyses suggest that the $e_g$ electrons on Mn sites occupy $d_{x^2-y^2}$ orbitals \cite{Mitchell97,kubota99a}, one may expect the metallic conductivity for the A-type AFM phase in La$_{2-2x}$Sr$_{1+2x}$Mn$_{2}$O$_{7}$ as it was the case for the Nd$_{1-x}$Sr$_{x}$MnO$_{3}$ system.\cite{kaw97,kuwahara99,mor98}  In contrast, no clear cusp nor other indication is present in the $T$ dependence of the resistivity (Fig. 1(c)).\cite{morR}

Concerning this question, an interesting observation was made in a recent X-ray and neutron diffraction study.  Very recently, Vasiliu-Doloc {\it et al.} have observed diffuse scattering in the paramagnetic phase of the $x=0.40$ sample which is possibly originated from a charge ordering (CO)\cite{Doloc99}.  If there exists the CO in the A-type AFM La$_{2-2x}$Sr$_{1+2x}$Mn$_{2}$O$_{7}$, it can naturally explain a lack of the metallic behavior of resistivity in the A-type AFM state.  Stimulated by this work, we have carried out systematic neutron diffraction studies on the 2D bilayer system La$_{2-2x}$Sr$_{1+2x}$Mn$_{2}$O$_{7}$ with $0.30 \le x \le 0.50$.  In the following, we shall demonstrate that the similar diffuse scattering exists for a surprisingly wide concentration of $0.30 \le x \le 0.50$ in La$_{2-2x}$Sr$_{1+2x}$Mn$_{2}$O$_{7}$.  We argue that it is consistent with short ranged stripe-like order, and causes the insulating behavior of the resistivity in the paramagnetic as well as A-type AFM state.

In the present study, the measurements were made on the same single crystal samples used in our previous works.\cite{hirota98,kubota99a}  Neutron diffraction measurements were carried out on triple-axis spectrometers GPTAS, TOPAN, and KSD in the JRR-3M of JAERI, Tokai, Japan.  Several incident neutron momentums were utilized, depending on the necessity of the intensity and the resolution, along with various combinations of collimators with PG filters.  The $(h 0 l)$ and $(h k 0)$ reciprocal planes were aligned as the scattering planes, and the samples were set in aluminum capsules filled with helium gas and were attached to the cold head of a closed-cycle helium gas refrigerator.

In order to examine a possible charge ordering (CO) within the MnO$_2$ planes in La$_{2-2x}$Sr$_{1+2x}$Mn$_{2}$O$_{7}$, we first searched for diffuse scattering in the $(h0l)$ zone, and observed a clear signal in all the samples we studied whose hole concentration ranges from $x=0.30$ through 0.50.  Figure 2 shows typical profiles of the diffuse scattering observed in the paramagnetic, A-type AFM, and canted AFM (FM) phase for the $x=0.40, 0.45$ and 0.48 samples, respectively.  The scans were made along the $[1 0 0]$ direction, and the diffuse scattering was observed around $Q = (2 \pm q_{o}, 0, 1)$.  The peak position of the diffuse scattering is located near $q_{o} \sim 0.3a^{*}$ for given three concentrations, but $q_{o}$ is slightly smaller and close to $\sim 0.25a^{*}$ for $0.30 \leq x < 0.40$ (not shown).

The diffuse scattering is further characterized by observing the profiles in the $(h k 0)$ scattering plane.  Scans was made around the two fundamental Bragg reflections $Q = (2 0 0)_{\rm N}$ and $(0 2 0)_{\rm N}$.  One may expect four satellites around each fundamental reflections which correspond to the modulation vector along the [100] or [010] direction.  However, only two satellites were observed along the longitudinal direction.  As shown in the upper panel of Fig. \ref{hk-scans_prof}, the diffuse scattering was observed around $Q = (2 \pm q_{o}, 0, 0)$ (longitudinal direction), but no intensity around $Q = (2, \pm q_{o}, 0)$ with $q_{o} \sim 0.3a^{*}$, respectively.  There was also no intensity at $Q = (2 \pm q_{o}, \pm q_{o}, 0)$.  More importantly, the width (inverse correlation length) of these diffuse peak is extremely anisotropic.  Comparing the profiles in Fig. \ref{hk-scans_prof}, one finds that the inverse correlation length along the longitudinal direction $\kappa_{\rm L} \sim 0.02a^{*}$, while $\kappa_{\rm T}$ perpendicular to the modulation is $\sim 0.2a^{*}$, being an order of magnitude larger than  $\kappa_{\rm L}$.

Vasiliu-Doloc {\it et al.} reported that the diffuse scattering in the x=0.40 FM sample is of nuclear origin.\cite{Doloc99}  In a strict terminology, a nuclear-origin satellite observed by neutron diffraction indicate the periodic lattice modulations.  From the beautiful correlation between the diffuse intensity and the resistivity observed in the present study shown in Fig. 5, it may be reasonable to interpret that the underlying charge ordering (CO) gives rise to the diffuse peaks.  Consequently, the present results have very important implication on the nature of the CO in the 2D bilayer manganite La$_{2-2x}$Sr$_{1+2x}$Mn$_{2}$O$_{7}$.  The results presented in Fig. \ref{h-scan_prof} and in Fig. \ref{hk-scans_prof} suggest that the CO is consistent with a stripe-type charge modulation along the $[1 0 0]$ direction ({\it parallel} to the MnO$_2$ bond direction), although its correlation length perpendicular to the stripe modulation is very short as is evident from the strong anisotropy between $\kappa_{\rm L}$ and $\kappa_{\rm T}$.  Let us compare the present {\it quasi-stripe CO} with the stripe order observed in the 2D Ni and Cu oxides.\cite{tra,kazuyamada}  The most noticeable difference is that the present quasi-stripe CO in the 2D bilayer manganite is formed within the FM MnO$_2$ planes.  On the other hand,  the stripes in the Ni and Cu compounds are formed in the matrices of the AFM spin order, and they act as the antiphase domain boundaries for the spin order.  It should be also pointed out that the incommensurability $q_{o} \sim 0.3a^{*}$ of La$_{2-2x}$Sr$_{1+2x}$Mn$_{2}$O$_{7}$ is almost independent of $x$ for $0.40 \le x \le 0.50$, but approaches $\sim 0.25a^{*}$ for $0.30 \leq x <0.40$.  This is remarkable contrast with the linearity between the incommensurability and the hole concentration observed in the underdoped cuprate\cite{kazuyamada} and the nickelate system, but resembles the overdoped cuprate superconductor.\cite{tra,kazuyamada,yoshizawa}

Now that the feature of the diffuse scattering within the MnO$_2$ layers has been well characterized, their stacking along the $c$ direction was examined by studying the profiles of the diffuse scattering along the $l$ direction on the $(h0l)$ scattering plane.  Reflecting the 2D character of the La$_{2-2x}$Sr$_{1+2x}$Mn$_{2}$O$_{7}$ system, very unusual profiles of the diffuse intensity were observed along the $l$ direction as depicted in Fig. \ref{l-scan_prof}.  The double-peaked local maxima were observed near $l \sim 0$ and $\sim 10$ associated with the $(h = 2n \pm 0.3, 0, l)$ lines.  In the inset, shaded area indicate the positions of the diffuse scattering, and filled circles highlight their peak positions.

We found that the observed characteristic profile along the $l$ direction was well described by the stacking of a stripe-like charge order.  Although the details of our analysis will be published elsewhere, we describe the outline and principal results of the analysis below.  We denote the structure factor of the quasi-stripe CO for the $i$-th MnO$_2$ bilayer and associated (La$_{1-x}$Sr$_x$)$_2$O$_2$ layers as $F_i$. The ($i+1$)-th block is separated from the $i$-th block by 1/2$c$ along the $c$ axis.  We define the strength of the correlation between adjacent layers as $r$.  Then, the correlation function $J_{m}$ of two blocks which are $m$-blocks apart may be expressed by
\begin{equation}
J_{m}=\langle F_{i} F_{i+m}\rangle=(-r)^{m}\langle F F\rangle,
\end{equation}
where $\langle \cdots \rangle$ means a thermal average of the correlation function and $\langle F F\rangle$ indicates an auto-correlation function.  Summing up  correlation functions  over all $m$, the intensity of the diffuse scattering due to the quasi-stripe order along the $l$ direction is given  as follows:
\begin{equation}
I(l)\propto \sum_{m=-\infty}^{\infty}J_m e^{-i\pi lm} = \frac{1-r^2}{1+r^2+2r\cos(\pi l)}\langle F F\rangle.
\label{l-str}
\end{equation}

In addition, by taking into account the distribution of the $z$ coordinates from the averaged coordinate $z_{o}$ for each atoms, we extended the $i$-th block structure factor $F_i$ .  In fact, the short-ranged quasi-stripe order causes the random deviation of individual $z$ coordinates from $z_{o}$.  The fit to eq. \ref{l-str} with the extended structure factor accounts well for the observed data as shown by a curve in Fig.\ref{l-scan_prof}.  The analysis also indicated that the stripes are paired within each bilayer, in other words, they are stacked in phase despite the repulsive Coulomb interaction.  Consequently, the CO in the bilayer La$_{2-2x}$Sr$_{1+2x}$Mn$_{2}$O$_{7}$ is characterized by a short ranged bi-stripe order.

Next, we argue the influence of the quasi-bistripe order to the resistivity.  A beautiful correspondence was observed between the $T$ dependence of the CO and that of the resistivity as shown in Fig. \ref{dif-and-pho_T-dep}.  The $x=0.40$ sample shows the A-type AFM phase below $T_{\rm N} \sim 150$ K, and the FM phase below $T_{\rm C}(\sim 120$ K).  The diffuse intensity of the $x=0.40$ sample grows as $T$ is lowered, but quickly disappears below $T_{\rm C}$, being consistent with the metallic behavior of the resistivity.

For the $x=0.45$ sample, the diffuse intensity shows a maximum near $T_{\rm C}$, and decreases in the canted AFM phase, but finite intensity remains at low temperatures.  The resistivity is insulating below 300 K, and monotonically increases down to $\sim T_{\rm C}$, then shows a slight decrease in the canted AFM phase.  Note that the resistivity of the $x=0.45$ sample at low temperatures is two orders of magnitude larger than that of the metallic phase of the $x=0.40$ sample due to the remnant charge ordering within the FM MnO$_2$ layers. 

The $x=0.48$ sample shows only the A-type AFM phase below $T_{\rm N} \sim 200$~K\cite{hirota98,kubota99a}.  The strong diffuse intensity grows as $T$ is lowered from 300 K, but levels off in the A-type AFM phase.  The resistivity of the $x=0.48$ sample shows the insulating behavior at all temperature, although an increasing rate of the resistivity with decreasing $T$ is slightly suppressed near $T_{\rm N}$ due to the development of the A-type AFM long range order.

Figure \ref{dif-and-pho_T-dep} suggests that the quasi-bistripe order controls the transport property in La$_{2-2x}$Sr$_{1+2x}$Mn$_{2}$O$_{7}$.  The paramagnetic and even FM MnO$_2 $ layers in the 2D La$_{2-2x}$Sr$_{1+2x}$Mn$_{2}$O$_{7}$ are insulating due to the existence of the quasi-bistripe order.  In other words, the quasi-bistripe order is stable against the 2D short and long range FM spin correlations within the MnO$_2$ bilayers, but is instabilized by the formation of 3D FM long range order, indicating the importance of bulk ferromagnetism and/or dimensionality of the FM order for the stability of the quasi-bistripe order and metallic conductivity.  In this context, it should be noted that the correlation length of the quasi-bistripe order is very anisotropic within the MnO$_2 $ layers, being an order of magnitude longer along the stripe direction as shown in Fig. \ref{hk-scans_prof}, but its implication remains an open question.

In a recent Raman study for La$_{1.2}$Sr$_{1.8}$Mn$_{2}$O$_{7}$, the tetragonal-symmetry-forbidden phonon modes are observed, and it was suggested that the dynamic CE-type charge/orbital order could be the origin of such phonon modes.\cite{yama99}  From the present study, we suggest that it can be originated from the quasi-bistripe order within the MnO$_2$ layers.

This work was supported by a Grant-In-Aid for Scientific Research from the Ministry of Education, Science and Culture, Japan.

%\end{thebibliography}

\figure{Temperature dependence of (a) A-type antiferromagnetic Bragg reflection (0 0 3), (b) inverse magnetization $H/M$ along the $ab$-plane and $c$ direction measured at 0.1 T, and (c)  resistivity in the $x=0.45$ sample.  Lines are guides to the eye.\label{R&H/M_T-dep}}

\figure{Profiles of the diffuse scattering from charge ordering in three A-type AFM samples with $x=0.40, 0.45$ and 0.48. Horizontal bars indicate the instrumental resolution.  For the $x=0.48$ sample, there is a small temperature-independent background near $h \leq 2.3$.\label{h-scan_prof}}

\figure{Profiles of the diffuse scattering from charge ordering observed on the $(h,k,0)$ plane in the $x=0.45$ sample. Thick bars labeled `Res' indicate the instrumental resolution.  Lines are guides to the eye.\label{hk-scans_prof}}

\figure{Profiles of the diffuse scattering from the charge ordering observed along the $(00l)$ direction along the $(2.3, 0, l)$ line in the A-type AFM phase at 120 K.  Inset: $(h, 0, l)$ reciprocal plane of the La$_{2-2x}$Sr$_{1+2x}$Mn$_{2}$O$_{7}$ system.  The shaded regions indicate the region where the diffuse scattering from the charge ordering was observed, and filled circles indicate the local maxima of the diffuse scattering.\label{l-scan_prof}}

\figure{Temperature dependence of the diffuse intensity observed at the peak position of the profiles shown in Fig. 2 (Left) and that of the resistivity (Right) for the $x=0.40, 0.45$ and 0.48 samples.\label{dif-and-pho_T-dep}}


\begin{references}
%\begin{thebibliography}{99}
\bibitem[*]{byline}
Present address: Photon Factory, Institute of Materials Structure Science,
KEK, 1-1 Oho, Tsukuba-shi, 305-0801,
E-mail: mkubota@post.kek.jp. \\Submitted to J. Phys. Soc. Jpn.

\bibitem{kaw97}
H. Kawano, R. Kajimoto, H. Yoshizawa, Y. Tomioka, H. Kuwahara, and Y. Tokura:
 Phys. Rev. Lett. {\bf 78}, 4253 (1997); 
H. Yoshizawa, H. Kawano, J. A. Fernandez-Baca, H. Kuwahara, and Y. Tokura: 
Phys. Rev. B {\bf 58},  (1998) R571.

\bibitem{kuwahara99}
H. Kuwahara, T. Okuda, Y. Tomioka, A. Asamitsu, and Y. Tokura:
Phys. Rev. Lett. {\bf 82} (1999) 4316.

\bibitem{mor98}
Y. Moritomo, T. Akimoto, A. Nakamura, K. Ohoyama, and M. Ohashi
Phys. Rev. B {\bf 58},  (1998) 5544.

\bibitem{hirota98}
K. Hirota, Y. Moritomo, H. Fujioka, M. Kubota, H. Yoshizawa, and Y. Endoh:
J. Phys.\ Soc.\ Jpn.  {\bf 67} (1998) 3380.

\bibitem{kubota99a}
M. Kubota, H. Fujioka, K. Ohoyama, K. Hirota, Y. Moritomo, H. Yoshizawa, and Y. Endoh: J. Phys. Chem. Solids {\bf 60} (1999) 1161; M.Kubota, H. Fujioka, K. Hirota, K. Ohoyama, Y. Moritomo, H. Yoshizawa, and Y. Endoh: J. Phys.\ Soc.\ Jpn. {\bf  69} (2000) June issue.

\bibitem{kubota99b}
M. Kubota, H. Yoshizawa, Y. Moritomo, H. Fujioka, K. Hirota, and Y. Endoh:
J. Phys.\ Soc.\ Jpn. {\bf  68} (1999) 2202.

\bibitem{Mitchell97}J. F. Mitchell, D. N. Argyriou, J. D. Jorgensen, D. G. Hinks, 
C. D. Potter, and S. D. Bader: Phys.\ Rev.\ B {\bf 55} (1997) 63; {\it ibid}. {\bf 55} (1997) R11965; {\it ibid}. {\bf 57} (1998) 72.

\bibitem{kim97}T. Kimura, A. Asamitsu, Y. Tomioka, and Y. Tokura: Phys.\ Rev.\ Lett. {\bf 79} (1997) 3720.


\bibitem{morR} We noticed that analysis of the resistivity indicated a slight change of the activation energy below $T_{\rm N}$.

\bibitem{Doloc99}L. Vasiliu-Doloc, S. Rosenkranz, R. Osborn, S. K. Sinha, 
J. W. Lynn, J. Mesot, O. H. Seek, G. Preosti, A. J. Fedro, and J. F. Mitchell:
Phys. Rev. Lett. {\bf 83} (1999) 4393.

\bibitem{tra}
J. M. Tranquada, B. J. Sternlieb, J. D. Axe, Y. Nakamura, and S. Uchida:
Nature {\bf 375} (1995) 561; Phys. Rev. B {\bf 54} (1996) 7489; Phys. Rev. Lett. {\bf 78} (1997) 338.

\bibitem{kazuyamada}K. Yamada, C. H. Lee, K. Kurahashi, J. Wada, S. Wakimoto, S. Ueki, H. Kimura, Y. Endoh, S. Hosoya, G. Shirane, R. J. Birgeneau, M. Greven, M. A. Kastner, and Y. J. Kim:
 Phys. Rev. B {\bf 57} (1998) 6165.

\bibitem{yoshizawa}
T. Katsufuji, T. Tanabe, Y. Tokura, T. Kakeshita, R. Kajimoto, and H. Yoshizawa:
Phys. Rev. B 60 (1999) R5097;
H. Yoshizawa,T. Kakeshita, R. Kajimoto, T. Tanabe, T. Katsufuji, and Y. Tokura:
Phys. Rev. B {\bf 61} (2000) R854.


\bibitem{yama99}
K. Yamamoto, T. Kimura, T. Ishikawa, T. Katsufuji, and Y. Tokura:
J. Phys.\ Soc.\ Jpn.  {\bf 68} (1999) 2538.

%\bibitem{moreo99} A. Moreo, S. Yunoki, and E. Dagotto: 
%Phys.\ Rev.\ Lett. {\bf 83} (1999) 2773.

\end{references}
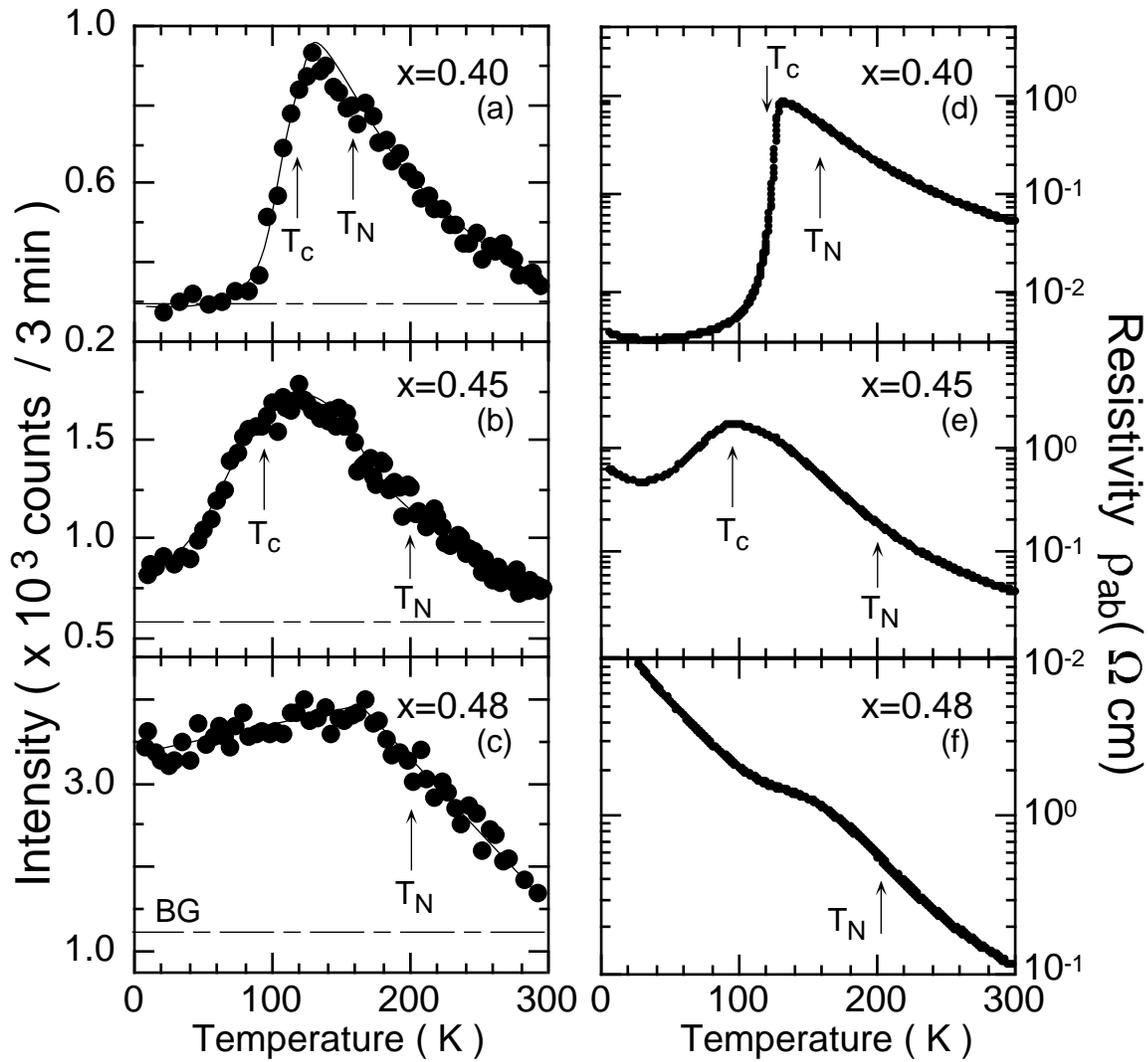
\end{document}